\documentclass[aps,prl,reprint,superscriptaddress,longbibliography]{revtex4-1}

\usepackage{graphicx}
\usepackage{amsmath}
\usepackage{epstopdf}
\usepackage{letltxmacro}
\usepackage[english]{babel}

\LetLtxMacro{\ORIGselectlanguage}{\selectlanguage}
\makeatletter
\DeclareRobustCommand{\selectlanguage}[1]{%
  \@ifundefined{alias@\string#1}
    {\ORIGselectlanguage{#1}}
    {\begingroup\edef\x{\endgroup
       \noexpand\ORIGselectlanguage{\@nameuse{alias@#1}}}\x}%
}
\newcommand{\definelanguagealias}[2]{%
  \@namedef{alias@#1}{#2}%
}
\makeatother

\definelanguagealias{en}{english}
\definelanguagealias{English}{english}

\begin{document}

\title{Tunnel-Junction Thermometry Down to Millikelvin Temperatures}

\author{A. V. Feshchenko} \email{anna.feshchenko@aalto.fi}
\affiliation{Low Temperature Laboratory, Department of Applied Physics, Aalto University, P. O. Box 13500, FI-00076 AALTO, Finland}
\author{L. Casparis}
\affiliation{Department of Physics, University of Basel, CH-4056 Basel, Switzerland}
\affiliation{Center for Quantum Devices, Niels Bohr Institute, University of Copenhagen, 2100 Copenhagen, Denmark}
\author{I. M. Khaymovich}
\affiliation{Low Temperature Laboratory, Department of Applied Physics, Aalto University, P. O. Box 13500, FI-00076 AALTO, Finland}
\author{D. Maradan}
\affiliation{Department of Physics, University of Basel, CH-4056 Basel, Switzerland}
\author{O.-P.~Saira}
\affiliation{Low Temperature Laboratory, Department of Applied Physics, Aalto University, P. O. Box 13500, FI-00076 AALTO, Finland}
\author{M. Palma}
\affiliation{Department of Physics, University of Basel, CH-4056 Basel, Switzerland}
\author{M. Meschke}
\affiliation{Low Temperature Laboratory, Department of Applied Physics, Aalto University, P. O. Box 13500, FI-00076 AALTO, Finland}
\author{J.~P.~Pekola}
\affiliation{Low Temperature Laboratory, Department of Applied Physics, Aalto University, P. O. Box 13500, FI-00076 AALTO, Finland}
\author{D.~M.~Zumb\"uhl}
\affiliation{Department of Physics, University of Basel, CH-4056 Basel, Switzerland}

\date{\today}

\begin{abstract}
We present a simple on-chip electronic thermometer with the potential to operate down to 1~mK. It is based on transport through a single normal-metal - superconductor tunnel junction with rapidly widening leads. The current through the junction is determined by the temperature of the normal electrode that is efficiently thermalized to the phonon bath, and it is virtually insensitive to the temperature of the superconductor, even when the latter is relatively far from equilibrium. We demonstrate here the operation of the device down to 7~mK and present a systematic thermal analysis.
\end{abstract}

\maketitle

\section{I. INTRODUCTION}

On-chip electronic thermometry is an important part of modern research and commercial applications of nano-technology, and it has been studied already for several decades; see Ref. \cite{Giazotto2006} and references therein. Many of these thermometers are based on tunnel junctions or quantum dots \cite{Gasparinetti2011, Mavalankar2013, Maradan2014}. Temperature sensors based on normal (N) and superconducting (S) metal tunnel junctions are used in a wide range of experiments \cite{Nahum1993, Nahum1994, Feshchenko2014} and applications \cite{Clark2004, Miller2008}. An example of such a device is a primary Coulomb blockade thermometer (CBT) that is based on normal-metal tunnel junctions with an insulator '$I$' as a tunnel barrier (NIN) \cite{Pekola1994, Casparis2012}, where the electronic temperature can be obtained by measuring the smearing of the single-electron blockade. One more example is a SNS thermometer \cite{Dubos2001}, whose critical current $I_{c}$ depends strongly on the temperature. Primary electronic thermometry has also been successfully demonstrated down to 10~mK using shot noise of a tunnel junction (SNT) \cite{Spietz2003, Spietz2006, Spietz2006a}. Nowadays, a standard dilution refrigerator reaches a temperature of 5-10~mK, with a record of 1.75~mK \cite{Frossati1978, Cousins1999}. Nevertheless, a thermometer that has a modest structure and a simple but accurate temperature reading at sub-10-mK temperatures and does not require a complicated experimental setup is still missing. For this purpose, we present a normal-metal - insulator - superconductor (NIS) junction that is widely used both as a refrigerating element and a probe of the local electronic temperature in different experiments and applications \cite{Nahum1993, Nahum1994, Leivo1996, Fisher1999, Clark2004,  Miller2008, Muhonen2012, Feshchenko2014}. The possibility to use the NIS junction at sub-10-mK temperatures makes this thermometer suited for cryogenic applications at low temperatures. For instance, quantum information is a highly focused and rapidly developing field in modern physics. For many realizations, such as superconducting and quantum-dot qubits, one needs to define a set of quantum states at low temperature, that are well separated and well controlled and insensitive to noise and decoherence effects \cite{Stern2010}. Several experimental realizations of two-level systems \cite{Dial2013, Simon2007, Paik2011} suggest that decreasing temperature further will increase the coherence times as well as improve charge sensitivity. We think that our thermometer will be interesting for a community who is willing to discover new physics as well as improve already existing devices that require low temperatures for their proper functioning. The NIS thermometer is easy to operate compared to SNT \cite{Spietz2003, Spietz2006}, and its thermalization is quite straightforward compared to CBT \cite{Casparis2012} due to the single-junction configuration and can be combined on chip with other solid-state devices. A measurement of the NIS current-voltage ($I-V$) characteristic yields a primary temperature reading.

In this paper, we study both experimentally and theoretically an on-chip electronic thermometer based on a single NIS tunnel junction at sub-10-mK temperatures. We demonstrate the operation of the NIS thermometer down to 7.3~mK. In addition, we develop a thermal model that explains our measurement data and shows that self-heating effects remain negligible for temperatures down to 1~mK.

\section{II. THEORETICAL BACKGROUND}

Transport through a NIS junction has strong bias and temperature dependence. Near zero bias voltage, the current is suppressed due to the superconducting gap, $\Delta$ \cite{Tinkham1996}. When biased at voltage $V$, current depends on the temperature due to the broadening of the Fermi distribution $f_{N}(E) = (e^{E/k_{B}T_{N}}+1)^{-1}$ in the normal metal with the temperature $T_{N}$ and the Boltzmann constant $k_{B}$. The current can be expressed as \cite{Tinkham1996}
\begin{equation}
I = \frac{1}{2 e R_{T}}\int_{-\infty}^{+\infty}{dE n_{s}(E)[f_{N}(E-eV) - f_{N}(E + eV)]},
\label{Eq.1}
\end{equation}
where $R_{T}$ is the tunneling resistance of the junction, and $E$ is the energy relative to the chemical potential.

In the superconductor, the Bardeen-Cooper-Schrieffer density of states is smeared and typically described by the Dynes parameter $\gamma$ expressed as $n_{s}(E) = \left| \Re e (u/\sqrt{u^{2} - 1})\right|$ (see the Supplemental Material in Ref. \cite{Pekola2010}), where $u = E/ \Delta(T_{S}) + i \gamma$, and $T_{S}$ is the temperature of the superconductor. Possible origins of $\gamma$ include broadening of the quasiparticle energy levels due to finite lifetime \cite{Dynes1978}, Andreev current \cite{Courtois2008, Greibe2011}, as well as photon-assisted tunneling caused by high-frequency noise and black-body radiation \cite{Pekola2010}. The typical experimental range of $\gamma$ for Al-based \footnote{Tunnel junctions based on Nb, NbN or NbTiN have higher $\gamma$ values, usually up to 10$^{-2}$ \cite{Jung2009, Nevala2012, Chaudhuri2013, Geresdi2015}} tunnel junctions is 10$^{-4}$ to 10$^{-5}$ for a single NIS junction \cite{Pekola2010, Nguyen2014}, getting as low as 10$^{-7}$ in SINIS single-electron transistors with multistage shielding \cite{Saira2012}.

\begin{figure}[h!]
\includegraphics[width=0.45\textwidth]{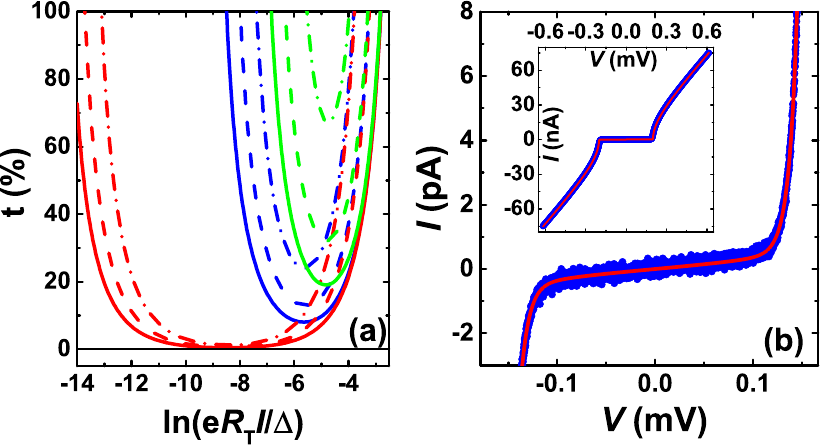}
\caption{In panel (a), we show the relative deviation of the present thermometer reading using method $\emph{B}$ with a numerically calculated $I-V$, Eq.~(\ref{Eq.1}). Sets of curves present the values of $t$ for three $\gamma$ parameters: 10$^{-7}$, 2.2$\times$10$^{-5}$ (the actual value in the experiment), and 10$^{-4}$ shown as the red, blue, and green curves from left to right. For each $\gamma$ parameter, temperatures 1, 3, and 7 mK are shown as dash-dotted, dashed, and solid lines, respectively. All curves are calculated using the parameters of the measured device with $\Delta$ = 200 $\mu$eV and $R_{T}$ = 7.7 k$\Omega$. The main panel (b) shows the measured $I-V$ characteristic (blue dots) together with full fit (solid red line) enlarged in the superconducting gap region. (Inset) Measured and calculated $I-V$ curve on a wider voltage scale at approximately 10 mK.}
\label{fig:1}
\end{figure}

One can determine $T_{N}$ from a measured $I-V$ curve using Eq.~(\ref{Eq.1}). As we show below, the self-heating of both N and S electrodes has a small effect on the $I-V$ characteristic; thus, for now, we neglect these effects. Therefore, we assume temperatures to be small,~$k_{B}T_{N, S}~\ll~eV,~\Delta(T_{S})$, and the superconducting gap to be constant  and equal to its zero-temperature value $\Delta$. In this case, for $eV < \Delta$, one can approximate Eq.~(\ref{Eq.1}) by
\begin{equation}
I \simeq I_{0} e^{-(\Delta - eV)/k_{B}T_{N}} + \frac{\gamma V}{R_{T} \sqrt{1-(eV/\Delta)^{2}}},
\label{Eq.2}
\end{equation}
where $I_{0} = \sqrt{2 \pi \Delta k_{B}T_{N}}/2e R_{T}$ \cite{Nahum1994, Courtois2008}. Here the second term stands for the corrections to the $I-V$ characteristic due to smearing. It leads to the saturation of the exponential increase of the current at low bias values. In the regime of moderate bias voltages, one can neglect this term and invert Eq.~(\ref{Eq.2}) into
\begin{equation}
V = \frac{\Delta}{e} + \frac{k_{B}T_{N}}{e} \ln(I/I_{0}).
\label{Eq.3}
\end{equation}
This equation provides a way to obtain the electronic temperature $T^{B}_{N}$ by only the fundamental constants and by the slope of the measured $I-V$ characteristic on a semilogarithmic scale as 
\begin{equation}
T^{B}_{N}(V) = \frac{e}{k_{B}}\frac{dV}{d(\ln I)}.
\label{Eq.4}
\end{equation}
This Eq.~(\ref{Eq.4}) allows us to use the NIS junction as a primary thermometer, however, with some limitations. One can include the effects of $\gamma$ into Eq.~(\ref{Eq.4}) by subtracting the last term in Eq.~(\ref{Eq.2}) from the current $I$ and obtain better accuracy. We do not take this approach here, since the main advantage of Eq.~(\ref{Eq.4}) is its simplicity as a primary thermometer without any fitting parameters.

Next, we will compare the two methods used to extract the electronic temperature from the measured $I-V$ curves. In method $\emph{A}$, we employ Eq.~(\ref{Eq.1}) and perform a nonlinear least-squares fit of a full $I-V$ characteristic with $T_{N}$ as the only free parameter. The value of $T_{N}$ obtained in this manner, named $T^{A}_{N}$, is not sensitive to~$\gamma$. Method $\emph{B}$ is based on the local slope of the $I-V$ [see Eq.~(\ref{Eq.4})]. The smearing parameter $\gamma$ has an influence on the slope of the $I-V$ characteristic and, thus, induces errors in the temperature determination. The temperature $T^{B}_{N}$ is extracted as the slope of measured $V$ vs $\ln I$ over a fixed $I$ range for all temperatures where Eq.~(\ref{Eq.4}) is valid. In the experiment, it is usually difficult to determine the environment parameters precisely, but one can determine $\gamma$ from the ratio of $R_{T}$ and the measured zero bias resistance of the junction. The $I-V$, which takes the $\gamma$ parameter into account [see Eq.~(\ref{Eq.2})], gives indistinguishable results from the ones obtained by method~$\emph{A}$. 

We evaluate the influence of the $\gamma$ parameter on the relative deviations of the present thermometer based on method $\emph{B}$ numerically, as shown in Fig.~\ref{fig:1}~(a). This deviation $t$ is defined as the relative error $t = (T^{B}_{N}/T_{N})-1$. We show the values of $t$ vs $\ln(e R_{T}I/\Delta)$ for two extreme cases $\gamma$ = 10$^{-7}$, 10$^{-4}$, and for $\gamma$ = 2.2$\times$10$^{-5}$ extracted from the present experiment at temperatures of 1, 3, and 7 mK. The lowest bath temperature is 3 mK, and 7 mK is the saturation of the electronic temperature in the current experiment. The larger the values of $\gamma$ and the lower the temperature, the higher the relative deviations. In addition, the range of the slope used to extract $T^{B}_{N}$ shrinks with increasing $\gamma$ and with decreasing temperature [see, e.g., red curves in Fig.~\ref{fig:1}~(a)]. Thus, reducing the leakage will significantly improve the accuracy of the device, especially towards lower temperatures. Possible avenues for suppressing $\gamma$ include improved shielding \cite{Saira2012, Hergenrother1995} and encapsulating the device between ground planes intended to reduce the influence of the electromagnetic environment \cite{Pekola2010}. Finally, higher tunneling resistance of the junction decreases the Andreev current \cite{Hekking1994}. We note that one can also use $dV$$/d(\ln g$) as a primary thermometer, where $g$ = $dI$/$dV$ is the differential conductance - typically a more precise measurement since it is done with a lock-in technique. Compared to Eq.~(\ref{Eq.4}), this method has the minimal deviation $t$ reduced by at least a factor of 3.5 for $T_{N}$ $\geq$ 1~mK (6 for $T_{N}$ $\geq$ 10~mK), though exhibiting qualitatively similar dependencies on $\gamma$ and $T_{N}$.

\section{III. EXPERIMENTAL REALIZATION AND MEASUREMENT TECHNIQUES}

\begin{figure}[h!t]
\includegraphics[width=0.45\textwidth]{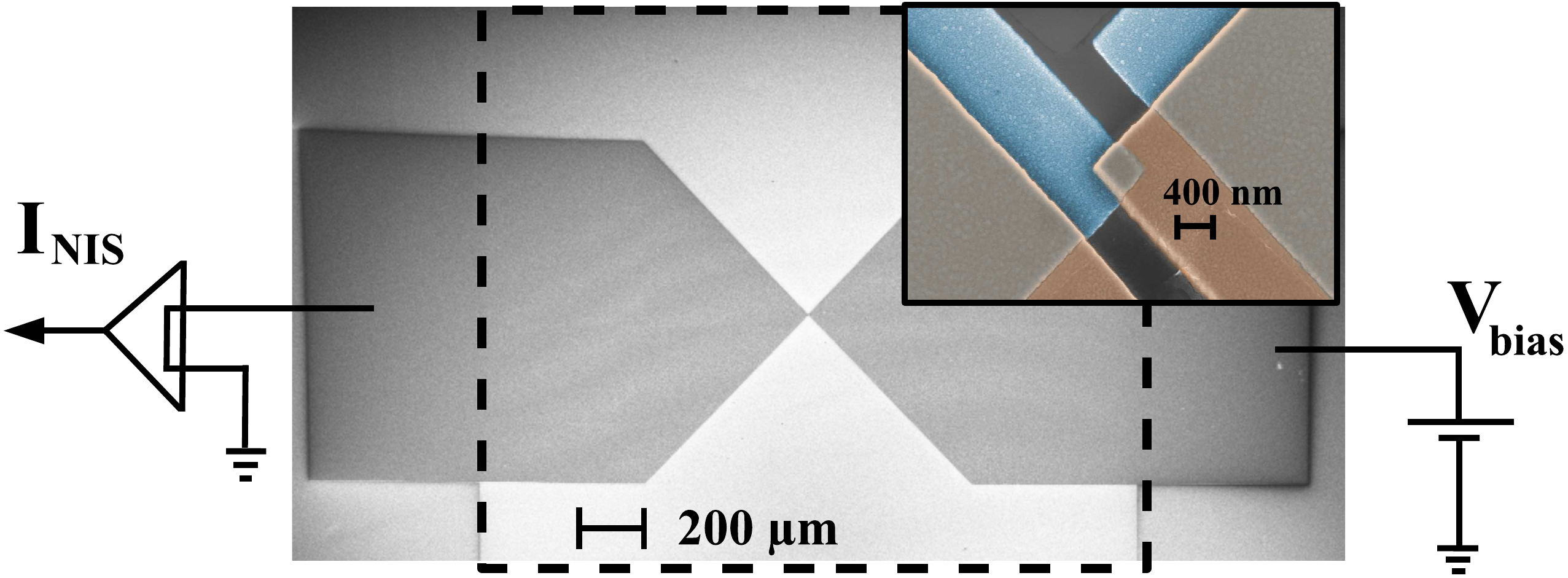}
\caption{A scanning-electron micrograph of the NIS device together with a schematic of the experimental setup. In the main panel, the S and N leads of the junction are visible, and underneath the pads, the ground plane of a square shape is indicated by a dashed line. Enlarged inset shows the actual tunnel junction where the S and N leads are shown in blue and brown, respectively.}
\label{fig:2}
\end{figure}

Next, we describe the realization of the NIS thermometer that is shown together with a schematic of the experimental setup in the scanning-electron micrograph in Fig.~\ref{fig:2}. 
The device is made by electron-beam lithography and the two-angle shadow evaporation technique \cite{Fulton1987}. The ground plane under the junction is made out of 50 nm of Au. To electrically isolate the ground plane from the junction, we cover the Au layer with 100 nm of AlO$_{x}$ using atomic-layer deposition. Next, we deposit a layer of $d_{S}$ = 40 nm of Al that is thermally oxidized \textit{in situ}. The last layer is formed immediately after the oxidation process by deposition of $d_{N}$ = 150 nm of Cu, thus, creating a NIS tunnel junction with an area $A$ = 380 nm $\times$ 400 nm. The geometry of the junction is chosen such that the leads immediately open up at an angle of 90$^{\circ}$ and create large pads with an area $A_{N} = A_{S}$ = 1.25 mm$^2$ providing good thermalization. The S lead is covered by a thick normal-metal shadow as shown in brown in the inset of Fig.~\ref{fig:2}, where the N and S layers are interfaced by the same insulating layer of AlO$_{x}$ as the junction.

The experiment is performed in a dilution refrigerator (base temperature 9~mK) where each of the sample wires is cooled by its own separate Cu nuclear refrigerator (NR) \cite{Clark2010}, here providing bath temperatures $T_{bath}$ down to 3~mK. Nuclear refrigerator temperatures after demagnetization are highly reproducible and obtained from the precooling temperatures and previously determined efficiencies \cite{Casparis2012}. Temperatures above approximately 9~mK are measured with a cerium magnesium nitrate thermometer which is calibrated against a standard superconducting fixed-point device. Since the sample is sensitive to the stray magnetic field of that applied on the nuclear refrigerator, this field is compensated down to below 1 G using a separate solenoid. The $I-V$ curves (see Fig.~\ref{fig:2} for the electrical circuit) are measured using a home-built current preamplifier with input offset-voltage stabilization \cite{Steinacher} to minimize distortions in the $I-V$ curves.

Filtering, radiation shielding, and thermalization are crucial for obtaining a low $\gamma$ and low device temperatures. Each sample wire goes through 1.6~m of thermocoax, followed by a silver epoxy microwave filter \cite{Scheller2014}, a 30-kHz low-pass filter, and a sintered silver heat exchanger in the mixing chamber before passing the Al heat switch and entering the Cu nuclear stage. The setup is described in detail in Ref. \cite{Casparis2012} and is further improved here (see the Appendix for more details).

\section{IV. RESULTS AND DISCUSSION}

In Fig.~\ref{fig:1}~(b), the measured $I$-$V$ characteristic in the superconducting gap region is shown by blue dots. The solid red line corresponds to the full fit based on method~$\emph{A}$. In the inset, we present the $I$-$V$ characteristic at a larger voltage scale, used to extract $R_{T}$.
\begin{figure}[h!t]
\includegraphics[width=0.4\textwidth]{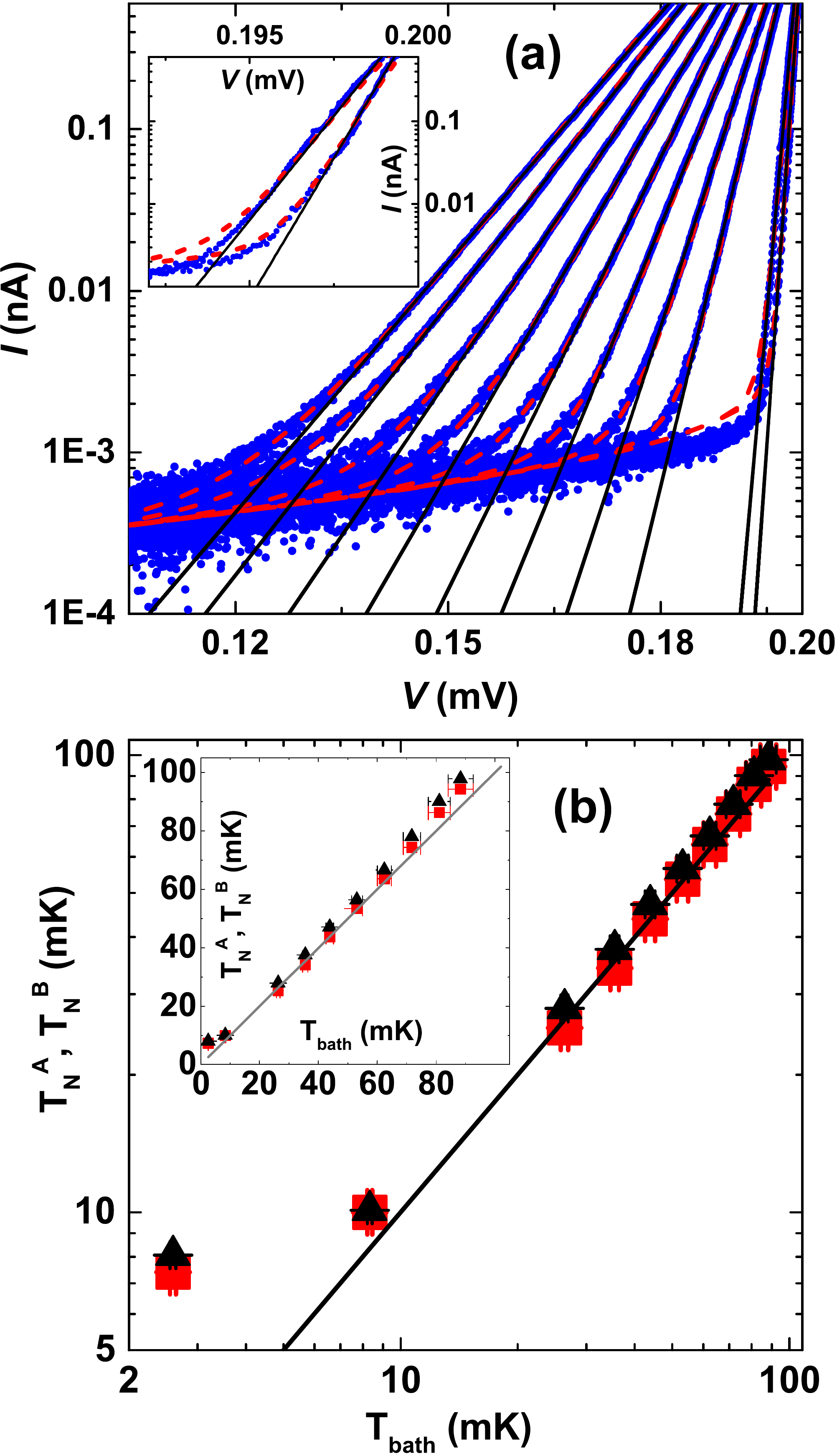}
\caption{Panel (a) shows measured $I$-$V$ characteristics (blue dots), when $T_{bath}$ is lowered from left to right together with fits as solid black and dashed red lines (see text). (Inset) Close-up of two $I$-$V$ characteristics for temperatures of 10 and 7 mK. The electronic temperatures extracted from both the full fit (red squares) of the $I$-$V$ characteristics and their slopes (black triangles) are shown in (b).}
\label{fig:3}
\end{figure}
In Fig.~\ref{fig:3}~(a), the measured $I$-$V$ characteristics of the NIS junction are shown in logarithmic scale by blue dots at $T_{bath}$ = $100, \ldots, 3$~mK from left to right. The full fits are shown as dashed red lines. The tunneling resistance $R_T$ = 7.7~k$\Omega$ and the Dynes parameter $\gamma$ = 2.2$\times$10$^{-5}$ used in all these fits are determined based on the $I$-$V$ characteristics shown in Fig.~\ref{fig:1}~(b) at high and low voltages, respectively. For the lowest temperatures, $T_{N}$ from the nonlinear fit depends strongly on the superconducting gap \footnote{A reduction of the superconducting gap by 0.1 $\%$ changes $T_{N}$ by 10 $\%$ at the lowest temperature.}, making it difficult to determine the gap with high enough accuracy \footnote{With the given experimental uncertainties, we can determine the gap with a precision of 0.25 $\%$ ($\pm$ 0.5 $\mu$eV).}. However, Eq.~(\ref{Eq.1}) gives a possibility to perform a nonlinear least-squares fit, and Eq.~(\ref{Eq.3}) gives a linear fit, where the parameters $\Delta$ and $T_{N}$ are responsible for the offset and the slope, respectively. Therefore, at high temperatures (approximately 100 mK in the present experiment), one can narrow down the uncertainty in $\Delta$ such that $T_{N}$ becomes essentially an independent parameter for the fits. Thus, the gap extracted from the high-temperature data using Eq.~(\ref{Eq.1}) remains fixed, $\Delta$ = 200 $\pm$ 0.5 $\mu$eV, for all temperatures below 100~mK. In addition, we show as solid black lines an exponential $I-V$ dependence corresponding to method $\emph{B}$ with a fitting range between 5 pA and 400 pA. The enlarged inset shows the $I-V$ curves at temperatures of 10 and 7 mK. The $I$-$V$ characteristics presented in Fig.~\ref{fig:3}~(a) agree well with the theoretical expressions in Eqs.~(\ref{Eq.1}) and (\ref{Eq.2}) (the latter form is not shown). In Fig.~\ref{fig:3}~(b), we show the electronic temperatures obtained using method $\emph{A}$ (red squares) and method $\emph{B}$ (black triangles) vs $T_{bath}$. Method $\emph{A}$ ($\emph{B}$) shows a relative error in the electronic temperature up to 6 $\%$ (11 $\%$). The error in method $\emph{B}$ is larger, as we neglect the influence of the $\gamma$ parameter. 

The lowest temperature obtained from the full fit is $T^{A}_{N}$~= 7.3 mK with statistical uncertainty of 5 $\%$ at $T_{bath}$ = 3 mK. The NIS temperature decreases slowly over time, arriving at 7.3~mK several weeks after the cooldown from room temperature. This suggests that internal relaxation causing a time-dependent heat leak, e.g., in the silver epoxy sample holder, is limiting the minimum temperature. Future improvements will employ low-heat-release materials better suited for ultralow temperatures such as sapphire or pure annealed metals, e.g., for the socket and chip carrier, minimizing organic noncrystalline substances such as epoxies.

\section{V. THERMAL MODEL}

The total power dissipated in the device is equal to $IV = \dot{Q}^{N}_{NIS} + \dot{Q}^{S}_{NIS}$, where $\dot{Q}^{N}_{NIS}$ and $\dot{Q}^{S}_{NIS}$ are the heat powers to the normal metal and to the superconductor, respectively. The heat released to the superconductor is given by
\begin{equation}
\dot{Q}^{S}_{NIS} = \frac{1}{e^{2}R_{T}}\int{E_{S} n_{s}(E)[f_{N}(E-eV) - f_{S}(E)]dE},
\label{Eq.5}
\end{equation}
where $E_{S} = E$ is the quasiparticle energy. To evaluate $\dot{Q}^{N}_{NIS}$, one has to substitute $E_{S}$ by $E_{N} = (eV-E)$ in Eq.~(\ref{Eq.5}). Almost all of the heat is delivered to the superconductor in the measured (subgap) bias range, thus, $\dot{Q}^{S}_{NIS} \sim IV$ and $\dot{Q}^{N}_{NIS}$ $\ll$ $\dot{Q}^{S}_{NIS}$.

So far, we neglected all self-heating effects both in the normal metal and in the superconductor. To justify the nonself-heating assumption, we check numerically and analytically these self-heating effects. We sketch the analytical arguments in the Appendix. Here we state the main results obtained from the thermal model.

Self-heating of the superconductor can take place due to the exponential suppression of thermal conductivity and the weak electron-phonon ($e-ph$) coupling, especially at low temperatures. We find that the superconductor temperature $T_{S}$ stays below 250 mK in the subgap bias range $\left| V\right| \le \frac{\Delta}{e}$ and does not influence the thermometer reading. In this bias range and at $T_{bath}$ = 3 mK, we estimate based on the numerical calculations the temperature of the superconductor $T_{S}$ = 145 mK and the power injected to the superconductor is $IV \sim 90$ fW. At the same time, we evaluate the relative change of the slope to be small $\left| t \right| \lesssim$ 5 $\times$ 10$^{-3}$ at $I \lesssim$ 1 nA. In conclusion, the temperatures obtained from both methods $\emph{A}$ and $\emph{B}$ are affected by less than 0.5 $\%$ by self-heating of the superconductor \footnote{The temperature of the superconductor would affect the $T_{N}$ through the dependence of the $I-V$ curve on the magnitude of the gap. The geometry of the device could influence the number of quasiparticles consequently the effective temperature of the superconductor.}. In addition, the normal metal might get self-heated due to weak electron-phonon coupling and backflow of heat from the superconductor \cite{Jochum1998}. The influence of the self-heating of the normal metal down to 1~mK temperature affects the temperature obtained from both methods $\emph{A}$ and $\emph{B}$ by less than 0.5 $\%$ as well, as in the case of self-heating of the superconductor.

\section{VI. CONCLUSIONS}

In conclusion, we demonstrate experimentally the operation of an electronic thermometer based on a single NIS tunnel junction. The thermometer agrees well with the refrigerator thermometer down to about 10~mK and reaches a lowest temperature of 7.3~mK at $T_{bath}$ = 3~mK, currently limited by a time-dependent heat leak to the sample stage. We discuss several possible improvements of the present device and experimental setup. Finally, we show that self-heating in the normal metal and in the superconductor on the full $I$-$V$ or its slope is negligible, paving the way for NIS thermometry down to 1~mK if the experimental challenges can be overcome.

\section{ACKNOWLEDGMENTS}
\begin{acknowledgments}
We thank G.~Frossati, G.~Pickett, V.~Shvarts, and M.~Steinacher for valuable input. We acknowledge the availability of the facilities and technical support by Otaniemi Research Infrastructure for Micro and Nanotechnologies. We acknowledge financial support from the European Community FP7 Marie Curie Initial Training Networks Action Q-NET 264034, MICROKELVIN (Project No. 228464), SOLID (Project No. 248629), INFERNOS (Project No. 308850), EMRP (Project No. SIB01-REG2), and the Academy of Finland (Project No. 284594 and No. 272218). This work is supported by Swiss Nanoscience Institute, NCCR QSIT, Swiss NSF, and ERC starting grant.

A. V. F. and L. C. contributed equally to this work.
\end{acknowledgments}

\appendix

\section{APPENDIX}

\subsection{A. Experimental techniques}

\begin{figure}[h!t]
\includegraphics[width=0.4\textwidth]{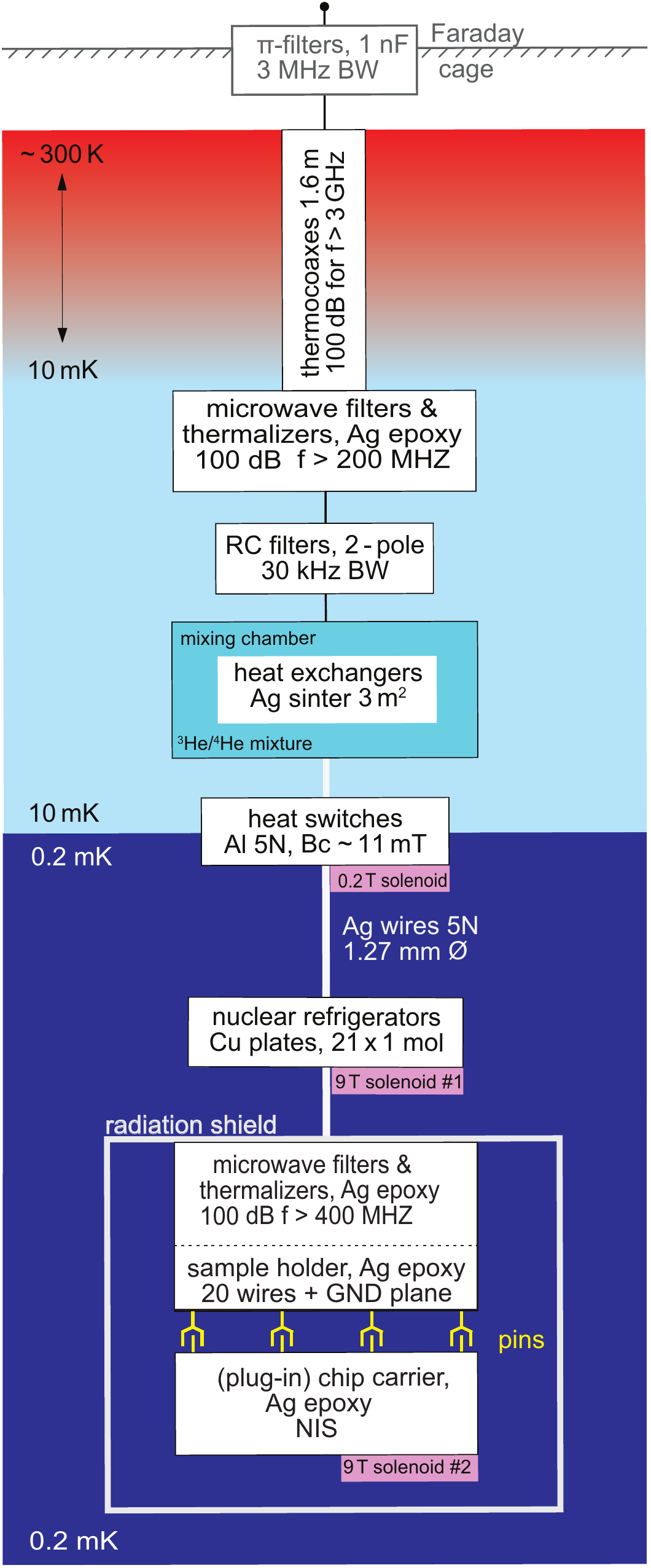}
\caption{Scheme of a dilution unit together with a NR. Radiation shields (not shown) are attached to the still and cold plate (approximately 50 mK). The resistor-capacitor (RC) filters are 1.6 k$\Omega$/2.2 nF and 2.4 k$\Omega$/470 pF. The 21 NR plates are 0.25 x 3.2 x 9.0 cm$^{3}$ each, amounting to 64-g Cu per plate. The NRs cool as low as 0.2 mK. In the present experiment, the lowest $T_{bath}$ used was ~3 mK. Compared to Ref. \cite{Casparis2012}, the improved setup depicted here features a Ag epoxy socket, a Ag epoxy chip carrier, and a second filtering stage with radiation-tight feedthroughs into an additional sample radiation shield. The abbreviations BW, B$_{c}$ and GND presented in the schematic stand for bandwidth, critical magnetic field and electrical ground, respectively.}
\label{fig:4}
\end{figure}

The setup described in Ref. \cite{Casparis2012} is improved as follows. First, a ceramic chip carrier is replaced by silver epoxy parts which remain metallic to the lowest temperatures, allowing more efficient cooling. Further, the sample -- previously mounted openly inside the cold-plate radiation shield together with the nuclear stages -- is enclosed in an additional silver shield, sealed with silver paint against the silver epoxy socket, and thermalized to one of the Cu refrigerators (see Fig.~\ref{fig:4}). Finally, each wire is fed into the sample shield through an additional silver epoxy microwave filter. While previously saturating at 10~mK or above \cite{Casparis2012}, metallic CBTs have given temperatures around 7~mK after the improvements \cite{Scheller2014, Scheller2014a}, comparable to the NIS temperatures presented here.

\subsection{B. Estimates of the subgap conductance}

The Dynes parameter $\gamma$ discussed in the main text can be attributed to the higher-order processes such as Andreev tunneling events. Assuming ballistic transport and an effective area of conduction channel $A_{ch} = 30~nm^{2}$ \cite{Maisi2011, Aref2011}, a simple estimate of subgap Andreev conductance reads $\sigma_{AR}$ = $R_{K}/(8NR_{T})$ = 8.5$\times 10^{-5}$ in units of $R_{T}^{-1}$, where $R_{K}$ is the quantum resistance, $N$ is the effective number of the conduction channels, and $N$ = $A$/$A_{ch}$, where $A$ is the area of the junction.  Alternatively, the estimate based on diffusively enhanced Andreev conductance yields of corresponding dimensionless conductance is 7.5$\times 10^{-5}$. These values are of the same order of magnitude as in our experiment ($\gamma$ = 2.2$\times 10^{-5}$) and fall in the range of earlier experiments \cite{Greibe2011}.

\subsection{C. Theoretical estimates for the relative deviations of the present thermometer}

\begin{figure}[h!t]
\includegraphics[width=0.30\textwidth]{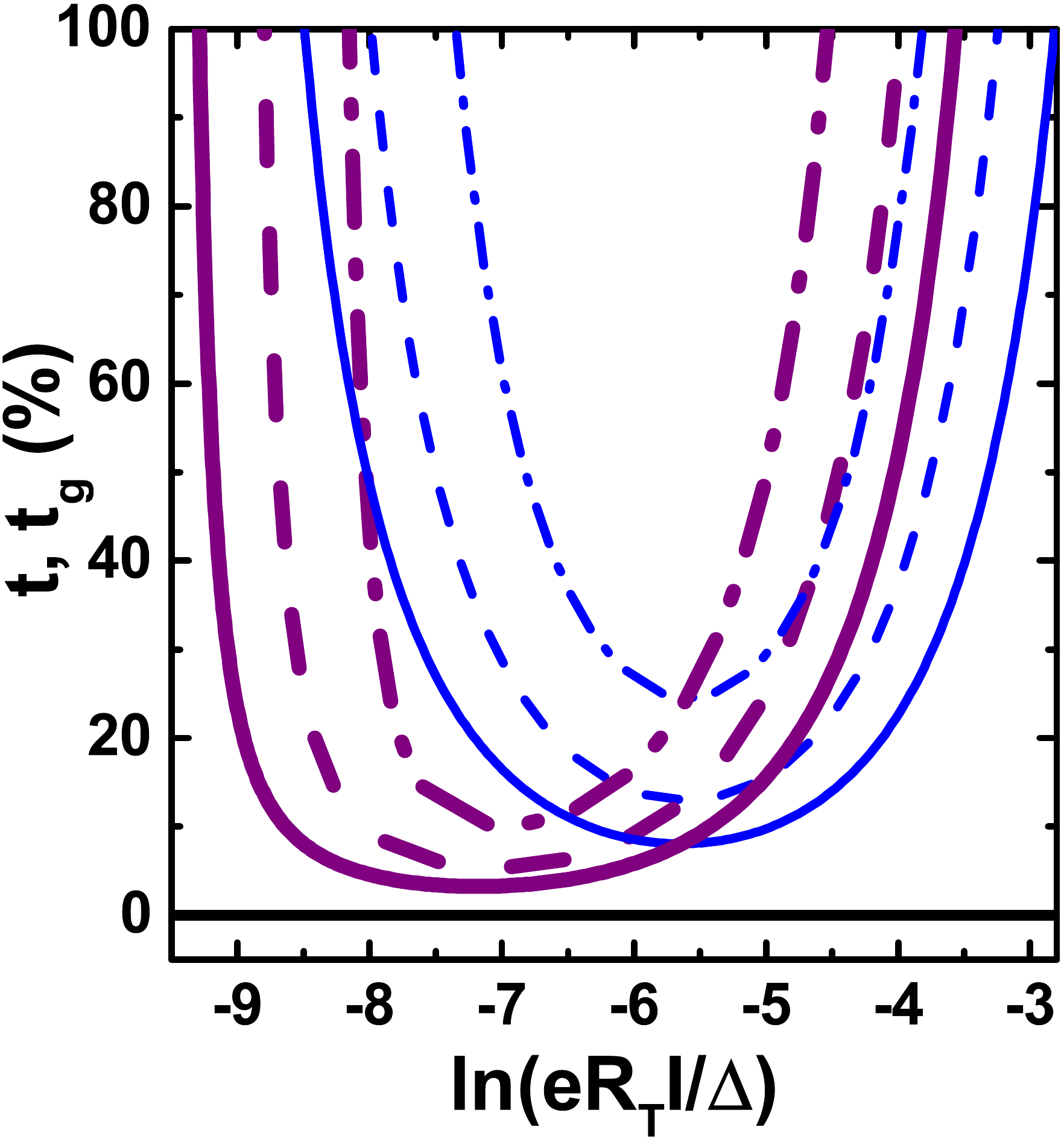}
\caption{The theoretical deviations $t$ and $t_g$ of the thermometer reading using method $\emph{B}$ based on $I(V)$ and $g(V)$, respectively. These deviations are shown for temperatures 1, 3, and 7 mK as dash-dotted, dashed, and solid lines, respectively. Parameters used are $\gamma$ = 2.2$\times$10$^{-5}$, $\Delta$ = 200 $\mu$eV, and $R_{T}$ = 7.7 k$\Omega$ are as in the actual experiment.}
\label{fig:5}
\end{figure}

The theoretical deviations of $dV/d(\ln I$) at~$\gamma$~=~2.2$\times 10^{-5}$ numerically calculated from $I-V$, Eq.~(\ref{Eq.1}), are rather large, particularly at low temperatures [approximately 30$\%$ at 1~mK; see solid blue curves in Figs.~\ref{fig:1}~(a) and \ref{fig:5}]. Measuring the differential conductance $g = dI/dV$ rather than current $I$ significantly reduces the predicted deviations $t_{g} = T_{N}^{slope, g}/T_{N}-1$, where $T_{N}^{slope, g} = \frac{dV}{d (\ln g)}e/k_{B}$. The minimum of these deviations gets broader and potentially reduces measurement noise since it is a lock-in measurement - overall strengthening the method $\emph{B}$. 
In Fig.~\ref{fig:5}, for comparison we show two sets of curves for $t_{g}$ (thick purple curves) and $t$ (thin blue curves) from left to right. Sets are calculated based on the experimental parameters for $\gamma$ = 2.2$\times$10$^{-5}$, $\Delta$ = 200 $\mu$eV and $R_{T}$ = 7.7 k$\Omega$. Each set corresponds to the temperatures 1, 3, and 7 mK and is shown as dash-dotted, dashed, and solid lines, respectively. Here, the $t$ set is identical to the set with $\gamma=2.2\times10^{-5}$ that is shown in Fig.~\ref{fig:1}~(a).

\subsection{D. Self-heating of the superconductor}

\begin{figure}[h!t]
\includegraphics[width=0.30\textwidth]{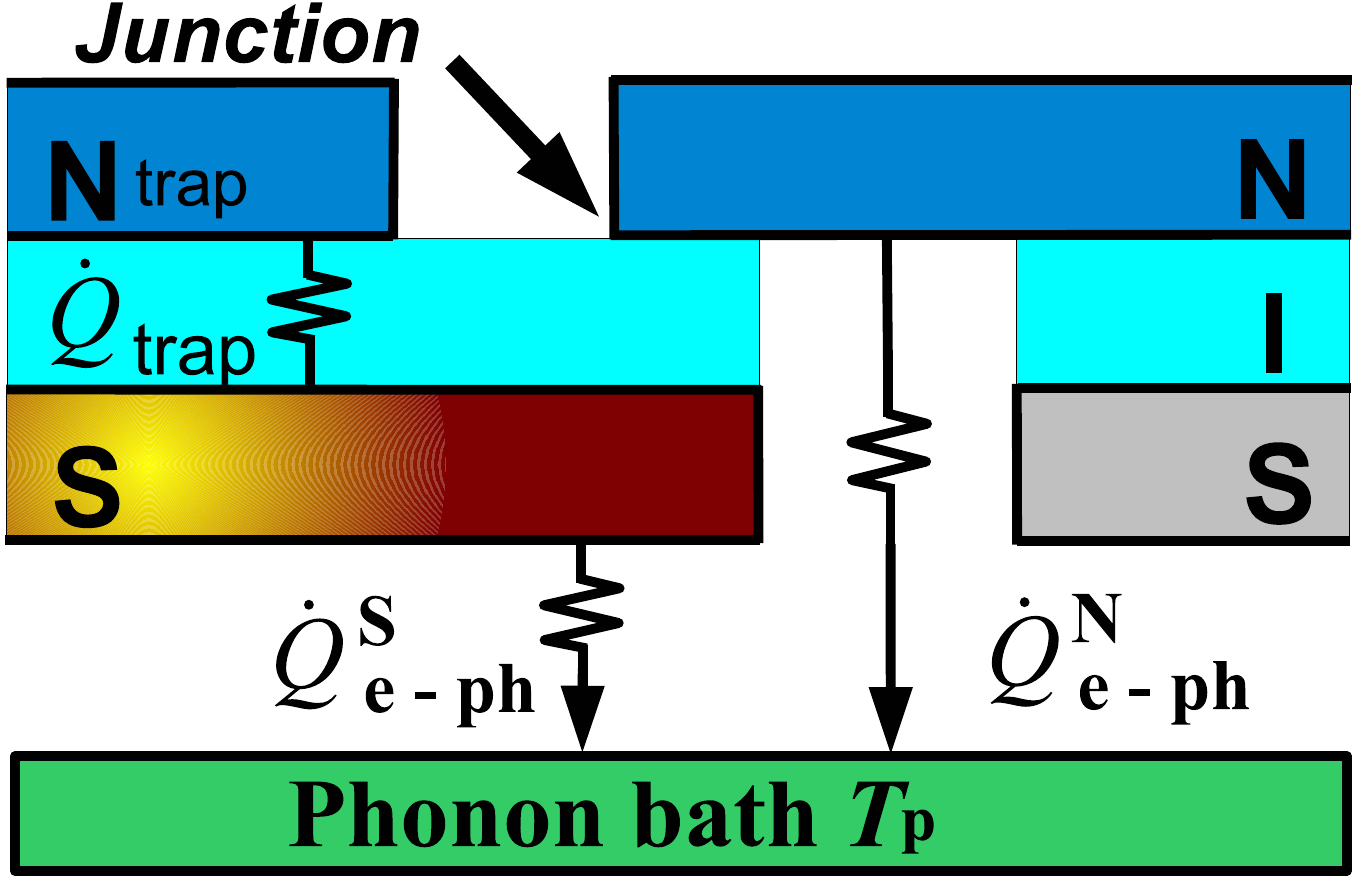}
\caption{The thermal diagram of the NIS thermometer. Present schematic does not reflect the real thicknesses of the materials. In this thermal model, we assume the normal-metal shadow that acts as the trap to be at $T_{bath}$.}
\label{fig:6}
\end{figure}

First, we consider the self-heating of the superconductor. We study the heat transport in the present geometry (see Fig.~\ref{fig:2}) by a diffusion equation assuming thermal quasiparticle energy distribution \cite{Visser2011, Knowles2012}
\begin{equation}
- \nabla (\kappa_{S}\nabla T_{S}) = u_{S},
\label{Eq.6}
\end{equation}
where we set a boundary condition near the junction $- \left.\bar{n}_{inner}\kappa_{S}\nabla T_{S}\right|_{junct} = \dot{Q}^{S}_{NIS}/A$, where $\bar{n}_{inner}$ is the inner normal to the junction. Thermal conductivity in the superconductor is
\begin{equation}
\kappa_{S} = \frac{6}{\pi^{2}}(\frac{\Delta}{k_{B}T_{S}})^{2} e^{\frac{-\Delta}{k_{B}T_{S}}} L_{0}T_{S}\sigma_{Al},
\label{Eq.7}
\end{equation}
where $L_{0}$ is the Lorenz number, and $\sigma_{Al}$ = 3 $\times$ 10$^{7}$ ($\Omega \mathrm{m})^{-1}$ is the electrical conductivity of the Al film in the normal state \cite{Knowles2012}, and we take into account the $T_{S}$ dependence of the gap at low temperatures, $\Delta (T_{S})/\Delta \simeq 1-\sqrt{2 \pi k_{B}T_{S}/\Delta}e^{-\Delta/k_{B}T_{S}}$. The absorbed heat is given by $u_{S}$ = $\dot{q}^{S}_{e-ph} + \dot{q}_{trap}$. Here, the first term is the electron-phonon power $\dot{q}^{S}_{e-ph} \simeq \Sigma_{Al} (T_{S}^{5} - T_{p}^{5}) \exp({-\Delta/k_{B}T_{S}})$ \cite{Timofeev2009}, where $\Sigma_{Al}$ = 3 $\times$ 10$^{8}$ WK$^{-5}$m$^{-3}$ is the material-dependent electron-phonon coupling constant. The phonon temperature $T_p$ is assumed to be equal to $T_{bath}$. Because of weak electron-phonon coupling, nearly all the heat is released through the (unbiased) normal-metal shadow (see Fig.~\ref{fig:6}) that acts as a trap for quasiparticles, $\dot{q}_{trap}$. 
Here the conductance of the trap per unit area is the same as for the tunnel junction $\sigma_{T} = 1/(R_{T}A)$. Therefore, the heat removed per volume by this trap $\dot{q}_{trap}$ can be calculated using Eq.~(\ref{Eq.5}) at $V$ = 0, $T_{N} = T_{bath}$ and substituting $R_{T}$ by $d_{S}/\sigma_{T}$. Thus, the temperature of the superconductor $T_{S}$ can be found as the solution of Eq.~(\ref{Eq.6}) in 2D in polar coordinates using radial approximation for the sample geometry and can be written as \cite{Knowles2012}
\begin{equation}
\sqrt{2\pi k_{B}T_{S}/\Delta(T_{S})} e^{-\Delta(T_{S})/k_{B}T_{S}} \equiv \alpha \dot{Q}_{NIS}^{S}.
\label{Eq.8}
\end{equation}
Here, we indeed assume $\dot{Q}_{NIS}^{S} \sim IV$, and $\alpha = \sqrt{\pi}e^{2}G/(d_{S}\sigma_{Al}\sqrt{2k_{B}T_{S}\Delta^{3}(T_{S})})$ is a coefficient that depends on $T_{S}$, and dimensionless parameter $G$~=~$\frac{\ln(\lambda/r_{0})}{\theta} \sim 2, \dots$, 3~is logarithmically dependent on the sample geometry \cite{Knowles2012}. Here, $\lambda$ is the relaxation length of the order of approximately 10, $\ldots$, 100 $\mu$m, and $r_{0} = 2A/(\pi d_{N})\simeq$ 500~nm is the radius of the contact in the present device. After substitution of all the parameters, we find that the superconductor temperature $T_{S}$  does not influence the thermometer reading staying below 250~mK in the subgap bias range $\left| V\right| \le \frac{\Delta}{e}$. We estimate $T_{S}$ to be approximately 145~mK in this bias range at $T_{bath}$ = 3~mK corresponding to the power injected to the superconductor as $IV \sim 90$ fW, and the quasiparticle density \cite{Knowles2012} as $n_{qp} = 0.3$ $\mu$m$^{-3}$. In addition, we evaluate the relative change of the slope to be small $\left| t \right| \lesssim$ 5 $\times$ 10$^{-3}$ at $I \lesssim$ 1 nA. In conclusion, the temperatures obtained from both methods $\emph{A}$ and $\emph{B}$ are affected by less than 0.5 $\%$ by self-heating of the superconductor.

\subsection{E. Self-heating of the normal metal}

For self-heating of the normal metal, one can solve the diffusion equation Eq.~(\ref{Eq.6}) taking into account the same boundary condition as above with all S indices replaced by N, where $\kappa_{N} = L_{0} \sigma_{Cu} T_{N}$ is the thermal conductivity of the normal metal, and the electrical conductivity of Cu is assumed to be $\sigma_{Cu}$ = 5 $\times$ 10$^{7}$  $(\Omega \mathrm{m})^{-1}$ \cite{Pekola2000}. The heat absorbed in the normal metal is $u_{N}$ = $\dot{q}^{N}_{e-ph} + \dot{q}^{N}_{wire}$. The heat conduction through the gold bonding wires $\dot{q}^{N}_{wire}$ is taken into account only at the point where it is attached to the normal-metal pad, whereas the electron-phonon interaction $\dot{q}^{N}_{e-ph}$ is effective in the full volume of the normal metal. The volumetric electron-phonon power is $\dot{q}^{N}_{e-ph} = \Sigma_{Cu} (T_{N}^{5} - T_{p}^{5})$, where $\Sigma_{Cu}$ = 2 $\times$ 10$^{9}$ W K$^{-5}$ m$^{-3}$ is the electron-phonon coupling constant of copper. Here we consider the effect of the heat removed by the bonding wires on the temperature only in the normal metal, thus, $\dot{q}^{N}_{wire} = L_{0}\sigma_{Au}(T_{N}^{2} - T_{bath}^{2})/2L_{wire}d_{N}$. The length of the gold bonding wire is $L_{wire} \simeq$ 5 mm and $\sigma_{Au}$ = 1.8 $\times$ 10$^{9}$ $(\Omega \mathrm{m})^{-1}$ is the electrical conductivity of gold measured at low temperatures. The thermal relaxation length in the normal metal \cite{Giazotto2006} is 
\begin{equation}
l_{N} = (T^{p/2-1}_{bath})^{-1}\sqrt{(\sigma_{Cu}L_{0}/(2\Sigma_{Cu})}.
\label{Eq.9}
\end{equation}
We substitute $p$ = 5 and $T_{bath}$ = 10 mK and obtain $l_{N}$ = 17.5 mm. Since all the dimensions of the present device are smaller than 1.5 mm, there is only a weak temperature gradient over the normal-metal electrode due to its good heat conduction, and the electron-phonon coupling weakens at low temperatures. By solving the heat-balance equation $\dot{Q}^{N}_{NIS}$ = $\dot{Q}^{N}_{e-ph} + \dot{Q}^{N}_{wire}$ and assuming no external heat leaks, one can calculate $T_{N}$. Here, the heat released through $e-ph$ coupling is $\dot{Q}^{N}_{e-ph} = \Omega_{N}\dot{q}^{N}_{e-ph}$, where $\Omega_{N} = A_{N}d_{N}$ is the volume of the N electrode. The heat released through $N_{wire}$ = 2 bonding wires is $\dot{Q}^{N}_{wire} = \dot{q}^{N}_{wire}N_{wire}A_{wire}d_{N}$, where its cross-sectional area is $A_{wire} = \pi r^{2}_{wire}$ with a radius $r_{wire}$ = 16 $\mu$m. The temperature obtained from both methods $\emph{A}$ and $\emph{B}$ is affected by less than 0.5 $\%$ by self-heating of the normal metal down to a temperature of 1~mK. In addition, we can evaluate at low temperatures $T_{N}$ $\leq$ $T_{S}$ $\ll$ $\Delta$ /$k_{B}$ the maximum cooling at optimum bias voltage $V_{opt}$ $\approx$ ($\Delta$  - 0.66 $k_{B} T_{N}$)/e \cite{Giazotto2006}
\begin{eqnarray}
\dot{Q}^{N}_{NIS}(V_{opt}) \approx \frac{\Delta^{2}}{e^{2}R_{T}}\left[ -0.59\left(\frac{k_{B}T_{N}}{\Delta}\right)^{3/2} + \right.\nonumber \\ \left. + \sqrt{\frac{2 \pi k_{B}T_{S}}{\Delta}}\times \exp{\left(-\frac{\Delta}{k_{B}T_{S}}\right)} + \gamma \right].
\label{Eq.10}
\end{eqnarray}
to be ~90 pW at $T_{bath}$ = 1~mK.

\end{document}